# High Infrared Reflectance Modulation in VO$_2$ Films Synthesized on Glass and ITO coated Glass substrates using Atmospheric Oxidation of Vanadium

Ashok P, Yogesh Singh Chauhan, and Amit Verma

Department of Electrical Engineering, Indian Institute of Technology Kanpur, Kanpur 208016, India

Email: ashok@iitk.ac.in and amitkver@iitk.ac.in

ABSTRACT: Vanadium Dioxide (VO$_2$) is a strongly correlated material, which exhibits insulator to metal transition at ~68 °C along with large resistivity and infrared optical reflectance modulation. In this work, we use atmospheric pressure thermal oxidation of Vanadium to synthesize VO$_2$ films on glass and ITO coated glass substrates. With the optimized short oxidation durations of 2 min and 4 min, the synthesized VO$_2$ film shows high optical reflectance switching in long-wavelength infrared (8-14 μm) on glass substrates and mid-wavelength infrared (3-5 μm) on ITO coated glass substrates, respectively. Peak reflectance switching values of ~76% (at 9 µm) and ~ 79% (at 4.6 µm) are obtained on the respective substrates, which are among the highest reported values. Using the reflectance data, we extract VO$_2$ complex refractive index in infrared wavelengths, in both the insulating and metallic phases. The extracted refractive index shows good agreement with VO$_2$ synthesized using other methods. This demonstration of high optical reflectance switching in VO$_2$ thin films, grown on low cost glass and ITO coated glass substrates, using a simple low thermal budget process will aid in enhancing VO$_2$ applications in the optical domain.

1. Introduction

VO$_2$ is a phase change material, which shows a sharp reversible insulator to metal transition around 68 °C, with upto five orders of change in resistivity [1]. Associated with this phase transition is a structural transition from insulating monoclinic phase to metallic rutile phase [2]. The resistivity jump in VO$_2$ is useful for electrical switching applications such as oscillators [3], memristors [4], radio-frequency switches [5], reconfigurable filters and antennas [6], steep switching field-effect transistors [7,8], and selectors for resistive random-access memory [9]. The phase transition in VO$_2$ also significantly changes its optical reflectance in infrared wavelengths [10,11]. This reflectance variation is useful for realizing various optical applications such as ultra-thin absorbers [11], tunable radiators [12,13], infrared camouflage [14, 15], microbolometers [16, 17], metamaterials [18, 19], and thermochromic films [20]. Most of these applications lie in mid-wavelength infrared (MWIR) (3-5 µm) and long-wavelength infrared (LWIR) (8-14 µm) part of the electromagnetic spectrum.

As multiple Vanadium oxide compounds are possible due to multiple valence states of Vanadium [21], precise control of oxygen partial pressure is needed for synthesizing high quality $VO_2$ thin films using traditional thin film deposition methods such as reactive sputtering [23-25], chemical vapor deposition [26-28], sol-gel synthesis [29], pulsed laser deposition [30, 31], reactive-evaporation [32], molecular beam epitaxy [33, 34], atomic layer deposition [35], and thermal oxidation of Vanadium [36, 37, 42]. Polymer assisted solution based process is an alternative method to synthesize $VO_2$ thin films without precise control of oxygen gas flow [38]. This method however requires high temperature (600 ˚C) annealing in nitrogen environment for long duration (3 hours) [38].

Atmospheric pressure thermal oxidation (APTO) is another synthesis method that does not need precise control of oxygen partial pressure as atmospheric oxygen is used to oxidize Vanadium thin films to synthesize $VO_2$ [39-44]. APTO method is relatively simple as the oxidation time and the oxidation temperature are the only parameters requiring precise control to synthesize $VO_2$ with good phase transition properties. Recently, we demonstrated low thermal budget synthesis of $VO_2$ with good electrical and optical switching properties on c-plane Sapphire substrates using a modified APTO method with a step oxidation temperature profile, which allows for precise control of oxygen content in the films [44]. As the oxidation duration increases, $VO_2$ content of the film increases till a certain optimum duration, leading to improved electrical and optical switching performance [42, 44]. After this optimum time, $V_2O_5$ content of the film starts to increase, leading to reduction in electrical and optical switching properties. For very long oxidation durations, phase pure $V_2O_5$ films are formed [42, 44]. Integration of $VO_2$ on low cost substrates such as glass and ITO/FTO/Al-ZnO coated glass is desired for widespread optical applications of $VO_2$. In this work, we apply the modified APTO method to synthesize $VO_2$ films on glass and ITO coated glass substrates, demonstrating among the highest reported LWIR/MWIR optical switching performance with an extremely low thermal budget process.

In section 2, we present the vanadium RF sputtering deposition conditions and the APTO process to synthesize the vanadium oxide films. In section 3, the optical characterization results of the samples using temperature-dependent Fourier-transform infrared spectroscopy (FTIR) reflectance measurements are presented to find the optical switching performance as a function of oxidation time. Temperature-dependent electrical resistance as a function of oxidation duration and surface morphology Atomic Force Microscopy (AFM) images are also presented. For samples with highest infrared reflectance switching, optical modeling of the reflectance data to extract the $VO_2$ complex refractive index in infrared wavelengths on glass and ITO coated glass substrates is also presented. Conclusions are presented in section 4.

2. **Experimental Procedure**

The complete experimental process to synthesize the vanadium oxide films is schematically shown in Fig. 1. Glass and ITO coated glass substrates (ITO thickness:150 nm, Sheet Resistance: 20 Ω/sq ) were first cleaned using acetone and isopropyl alcohol in a sonicator, and blow dried using dry nitrogen before loading into the sputtering chamber.

Prior to the deposition, the sputtering chamber was pumped to base pressure of 5 x$10^{-4}$ Pa. Vanadium thin films were deposited using RF magnetron sputtering from a 2-inch diameter Vanadium target of 99.9% purity. During the deposition, Argon gas flow rate was kept at 30 sccm using a mass flow controller to maintain the chamber pressure at 1 Pa and RF power to the sputtering gun was maintained at 90 W. Substrates were kept at room temperature during 90 minutes of V deposition. The deposited V film thickness was measured using KLA-Tencor stylus profilometer and was found to be ~130 nm.

The V deposited substrates were unloaded from the sputtering chamber and diced into smaller pieces to have a total of 10 V/glass samples and 7 V/ITO/glass samples. These samples were then oxidized in open atmosphere at 450 ˚C for different durations ($T_{oxd}$) upto 20 minutes followed by quenching on a cold plate (at room temperature). More details on the process can be obtained from our previous work [44]. After sample preparation, room temperature infrared optical reflectance of all the samples was measured by an IR microscope (Agilent, Cary 600) connected to the FTIR spectrometer (Agilent, Cary 660). $VO_2$ features low infrared reflectance in the insulating phase and high infrared reflectance in the metallic phase. The reflectance modulation was obtained by repeating the FTIR measurements with samples kept at 100 ˚C which is beyond the phase transition temperature of $VO_2$. Temperature-dependent electrical resistance was measured using four-point probe set up with a temperature controlled oven. Surface morphological characterization was done using AFM (Oxford Instruments MFP-3D).

3.  **Characterization and Modeling**

Atmospheric oxidation of V thin film proceeds with the formation of $VO_2$ in the initial oxidation phase and $V_2O_5$ in the latter stage. This oxidation behavior has been confirmed by previous APTO works [39,42,44]. The electrical and infrared optical switching properties of the vanadium oxide films maximizes for an optimum $T_{oxd}$ with high $VO_2$ content in the films [44]. Figure 2 shows the measured optical reflectance switching as a function of $T_{oxd}$ for the vanadium oxide on glass substrates at 9 µm wavelength (Fig. 2(a)) and on ITO coated glass substrates at 4.6 µm (Fig. 2(b)). Peak reflectance switching values of ~76% is obtained for $T_{oxd}$ = 2 minutes on glass substrate and ~79% for $T_{oxd}$ = 4 minutes on ITO coated glass substrate. Beyond these optimum oxidation times, optical reflectance switching decreases because of increasing $V_2O_5$ content of the films. The rate of decrease of optical reflectance modulation with increasing $T_{oxd}$ is observed to be much faster on glass substrate compared to ITO coated substrates.

Fig. 2(c) shows four-point probe resistance switching ratio ($R_{30\ ˚C}/R_{110\ ˚C}$) of the Vanadium oxide samples on glass substrate as a function of the oxidation duration $T_{oxd}$. The resistance switching first increases with increasing $T_{oxd}$, peaking at $T_{oxd}$ = 2 min, and decreases with further increase in $T_{oxd}$ because of increasing $V_2O_5$ content of the films. The resistance switching data and the infrared optical reflectance switching data are in agreement. Both characterizations suggest that $T_{oxd}$ = 2 min sample has the highest $VO_2$ content. Fig. 2(d) shows the resistance variation as a function of temperature for the $T_{oxd}$ = 2 min $VO_2$/glass sample, showing more than one order of resistance switching with phase transition temperature close to bulk $VO_2$ phase transition temperature of ~ 68 ºC. For vanadium oxide films on ITO substrates, we were not able to measure the electrical switching properties reliably as the highly conducting ITO layer provides a low resistance current pathway, dominating the electrical measurements.

AFM images (10µm x 10µm) for peak reflectance switching samples on glass and ITO coated glass substrate are shown in Fig. 2(e) and Fig. 2(f), respectively. Grain size is observed to be larger for the $VO_2$ sample on ITO coated glass substrate compared to the $VO_2$ sample on glass substrate. The surface RMS roughness is ~ 6.5 nm for $VO_2$/Glass sample and ~8.2 nm for $VO_2$/ITO sample.

The room temperature and high temperature (100 ˚C) FTIR reflectance data (4.5-14 µm) for the sample with optimum oxidation duration on glass substrate ($T_{oxd}$ = 2 minutes) is shown in Fig. 3(a). Large reflectance switching in LWIR (8-14 µm) wavelengths is clearly observed with maximum switching (~76%) observed at ~ 9 µm. To extract the complex refractive index of $VO_2$ from this data, we use Drude-Lorentz oscillator model to describe the optical properties of the film and substrates [22]. According to this model, the complex dielectric response of any material at frequency ω is given as:

$$\varepsilon(\omega) = \varepsilon_\infty - \frac{\omega_p^2}{\omega(\omega+i\omega_c)} + \sum_j \frac{f_j}{1-\frac{\omega^2}{\omega_j^2}-i\gamma_j\omega/\omega_j} \qquad (1)$$

where, the first term of $\varepsilon_\infty$ is the high frequency permittivity, the second term is Drude oscillator in which $\omega_p$ corresponds to plasma frequency and $\omega_c$ is the collision frequency. The third term is the Lorentz oscillator, where $\omega_j$ is the resonance frequency, $f_j$ is the strength of oscillator, and $\gamma_j$ is the line width.

We have calculated reflectance spectra using the Drude-Lorentz oscillator to extract the complex refractive index (n,k) of the $VO_2$ film in the insulating and the metallic states. To find the optimized oscillator parameters, we used constrained non-linear least-square fitting of the data. Three Lorentz oscillators are used to fit the measured reflectance spectra of $VO_2$ thin films grown on glass in the insulating phase, whereas, Drude oscillator alone is used to fit $VO_2$ reflectance in the metallic state. The optimized Lorentz and Drude oscillator parameters are summarized in table 1 and the calculated reflectance is shown in Fig. 3(a). Good fit is obtained with the measured reflectance data. The extracted n-k of $VO_2$ on glass is shown in Fig. 3(b) for the insulating phase and in Fig. 3(c) for the metallic phase. For comparison, n-k values of $VO_2$ deposited using sputtering and sol-gel methods from literature [45] is also shown. The extracted values of n and k match reasonably well with previous reports [45].

Fig. 4(a) shows the FTIR reflectance data (2.8-14 µm) at room temperature and at high temperature (100 ºC) for $T_{oxd}$ = 4 min sample on ITO coated glass substrate. Large reflectance switching in MWIR (3-5 µm) wavelengths is observed with maximum switching (~79%) observed at ~ 4.6 µm. Drude-Lorentz oscillator fit to the reflectance data is also shown in Fig. 4(a) and the optimized model parameters are summarized in table 1. In the $VO_2$ insulating phase, our fit exactly matches the measured reflectance spectra valley around 4.6 µm. For high-temperature metallic phase, fit matches reasonably well with the measured reflectance except for the small dip around 10 µm. The origin of this dip is not clear, it has been suggested that such a dip can be due to surface $V_2O_5$ layer which does not switch to metallic phase [45]. The extracted plasma frequency from Drude term agrees well with previous reports [22]. From this fitted reflectance spectra, we extracted n and k for both the insulating and the metallic phases which are shown in Fig. 4(b,c). The extracted (n,k) values match reasonably well with the reported (n,k) values of $VO_2$ films synthesized using other methods [45].

Table 2 summarizes the reported infrared reflectance switching of $VO_2$ synthesized using different methods and on different substrates. The synthesized $VO_2$ films in this work, using the modified APTO method, show among the highest reported optical reflectance switching performance.

## 4. Conclusion

In this work, we used a simple atmospheric pressure thermal oxidation method to synthesize optically switching $VO_2$ thin films on glass and ITO coated glass substrates. Highest optical switching of ~76% is obtained on glass substrate in LWIR and ~79% on ITO coated glass substrate in MWIR, with low thermal oxidation budgets of 450 ºC/2 minutes and 450 ºC/4 minutes, respectively. We also performed fitting of the reflectance data to extract the complex refractive indices of the $VO_2$ films which matches well with the optical properties of $VO_2$ synthesized using other methods. This demonstration of a simple $VO_2$ synthesis process with low thermal budget and low RMS roughness while giving extremely good optical switching performance on low cost substrates should expand the optical applications space of this phase transition material.

**Acknowledgements**

This project was supported by IIT Kanpur initiation grant. Ashok P would like to thank Dr.Jitendra Pradhan for useful discussions on reflectance modeling.

Table 1. The optimized Drude-Lorentz oscillator parameters of VO$_2$ films (Thickness 260 nm) on glass and ITO coated glass substrates in both the insulating and the metallic phases.

| | $\varepsilon_\infty$ | $\hbar\omega_p$ (eV) | $\hbar\omega_c$ (eV) | $\hbar\omega_1$ (eV) | $\hbar\omega_2$ (eV) | $\hbar\omega_3$ (eV) | $\gamma_1$ | $\gamma_2$ | $\gamma_3$ | $f_1$ | $f_2$ | $f_3$ |
|---|---|---|---|---|---|---|---|---|---|---|---|---|
| | | | | **Glass Substrate** | | | | | | | | |
| Insulating phase | 6.09 | - | - | 0.0762 | 0.025 | 1 | 2.177 | 0.1 | 0.1 | 3.89 | 194.4 | 8.578 |
| Metallic phase | 44.02 | 2.67 | 0.19 | - | - | - | - | - | - | - | - | - |
| | | | | **ITO coated Glass Substrate** | | | | | | | | |
| Insulating phase | 6.09 | - | - | 0.079 | 0.032 | 1.3 | 0.08 | 0.01 | 0.3605 | 1.9 | 48 | 4 |
| Metallic phase | 10 | 2.22 | 0.21 | - | - | - | - | - | - | - | - | - |

Table 2. Infrared peak reflectance modulation (ΔR %) of $VO_2$ synthesized using various methods and substrates.

| Synthesis Process | Substrate | $VO_2$ Thickness (nm) | IR Wavelength (μm) | ΔR % (Metallic R%- Insulating R%) |
|---|---|---|---|---|
| RF Sputtering [11] | c-plane sapphire | 180 | ~10 | ~75% |
| RF Sputtering [46] | c-plane sapphire | 50 | ~9.5 | ~52% |
| RF Sputtering [45] | Silicon | 130 | ~8 | ~50% |
| RF Sputtering [47] | Quartz | 200 | ~7.5 | ~42% |
| RF Sputtering [48] | Quartz | 300 | ~7.5 | ~48% |
| RF Sputtering [47] | ZnO | 100 | ~10 | ~70% |
| RF Sputtering [47] | AZO | 100 | ~5 | ~32% |
| RF Sputtering [48] | Al | 300 | ~8 | ~ -33% |
| RF Sputtering [49] | Silicon | 200 | ~3.5-4 | ~77% |
| DC Sputtering [50] | Glass | 130 | ~8.5-9 | ~75% |
| CVD [22] | ITO | 250 | ~4.8 | ~76% |
| CVD [22] | Glass | 253 | ~9 | ~75% |
| PLD [51] | $SiO_2$/ Si | - | ~8 | ~60% |
| PLD [52] | c-plane sapphire | 200 | ~10 | ~79% |
| PLD [53] | c-plane sapphire | 150 | ~10 | ~77% |
| PLD [52] | r-plane sapphire | 200 | ~10 | ~76% |
| PLD [54] | Pt-Si | 50 | ~3-4 | ~-20% |
| PLD [55] | Glass | 87 | ~7.5-8 | ~70% |
| Sol-gel [45] | Silicon | 110 | ~8 | ~45% |
| Sol-gel [50] | Glass | - | ~9-9.5 | ~60% |
| Sol-gel [50] | Glass | 700 | ~9-9.5 | ~50% |
| Decomposed from $V_2O_5$ Crystal [10] | Single crystal $VO_2$ | Bulk | ~17 | ~80% |
| APTO [44] | c-plane sapphire | 280 | ~9.6 | ~60% |
| **APTO (This Work)** | **Glass** | **260** | **~9** | **~76%** |
| **APTO (This Work)** | **ITO** | **260** | **~4.6** | **~79%** |

**FIGURES**

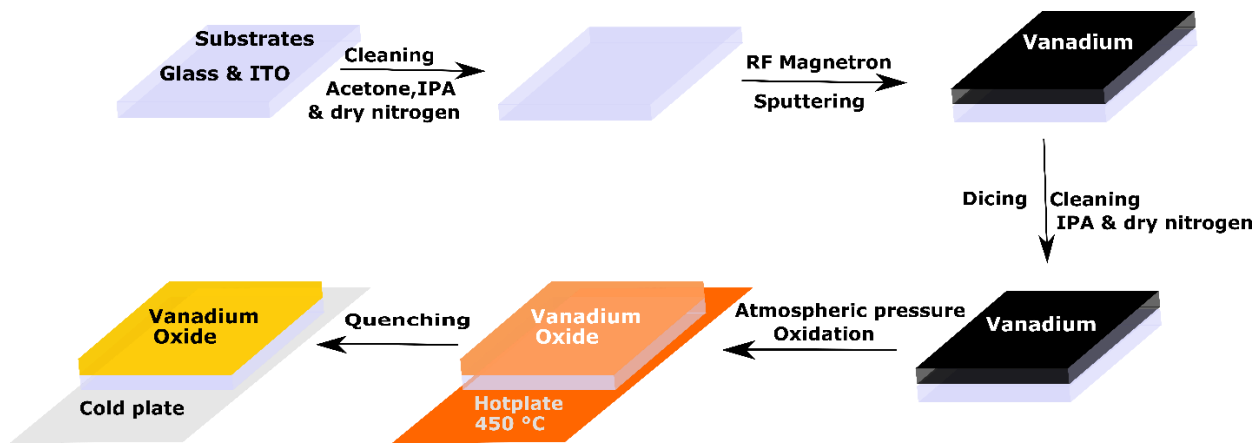

FIG. 1. Modified atmospheric pressure thermal oxidation process used to synthesize Vanadium oxide thin films.

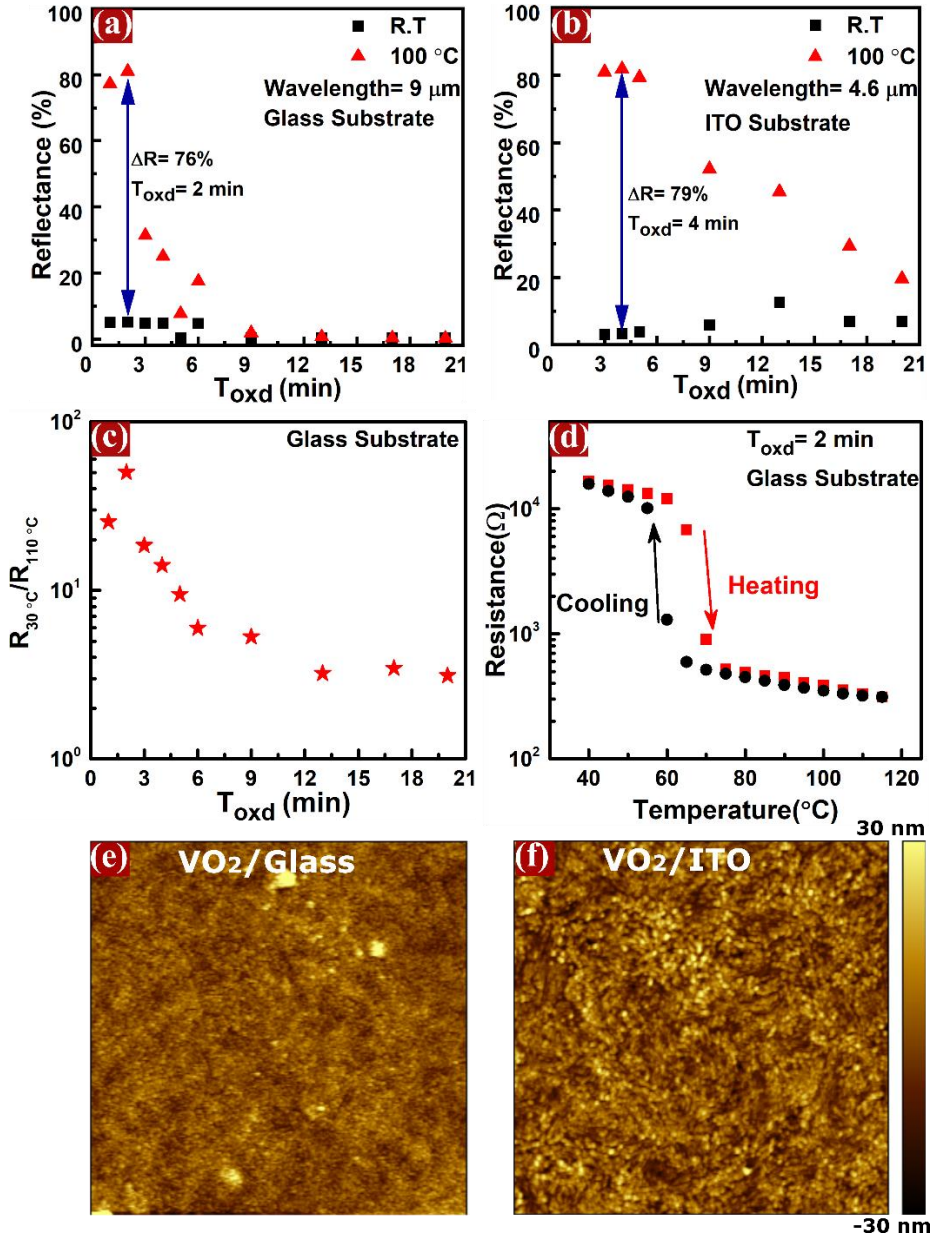

FIG. 2. Measured optical reflectance of the synthesized films as a function of the oxidation duration $T_{oxd}$, at room temperature and 100°C, at (a) 9 µm on glass substrates and (b) 4.6 µm on ITO coated glass substrates. Highest switching of ~76% is obtained for $T_{oxd}$ = 2 min sample on glass and highest switching of ~79% is observed for $T_{oxd}$ = 4 min oxidized sample on ITO coated glass substrate. (c) Four-point probe resistance switching ratio ($R_{30\,°C}/R_{110\,°C}$) as a function of $T_{oxd}$ for the Vanadium oxide samples on glass substrate. (d) Resistance of $T_{oxd}$ = 2 min sample as a function of temperature. AFM images (10 µm x 10 µm) of peak reflectance switching samples on (e) glass substrate and (f) ITO coated glass substrate.

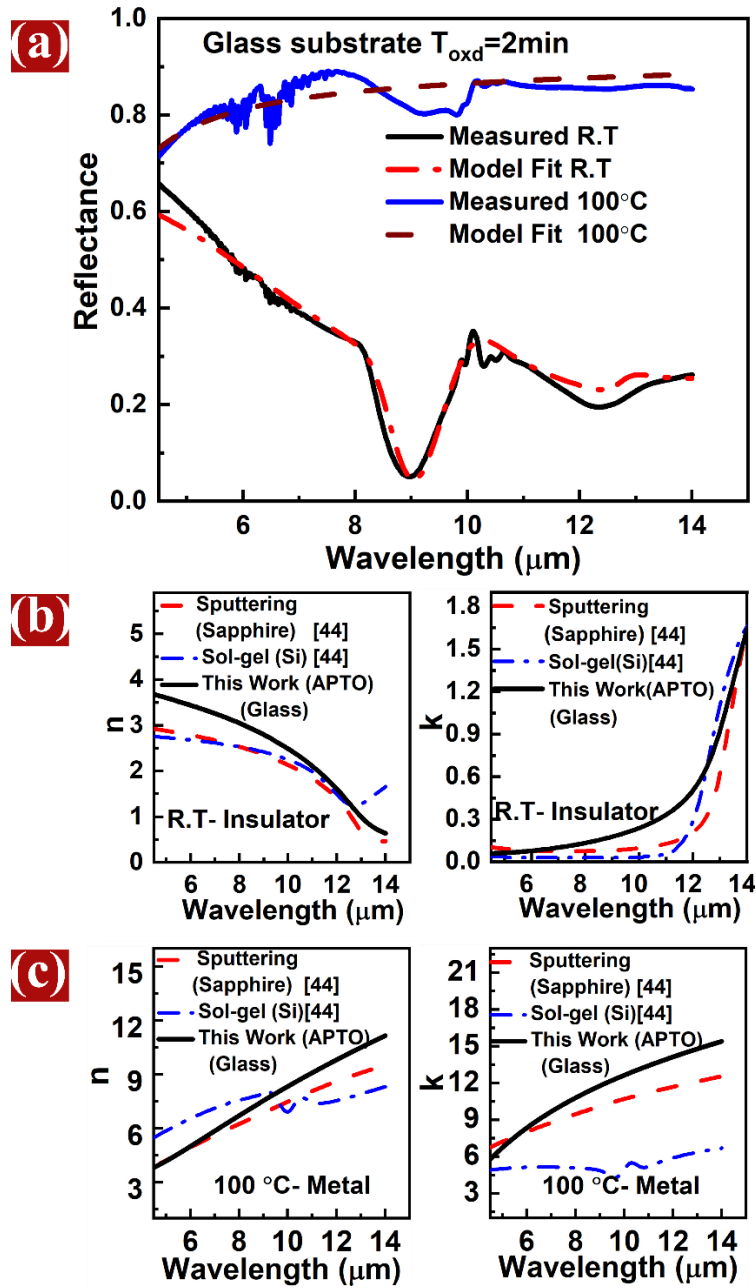

FIG. 3.(a) Measured infrared reflectance spectrum at room temperature and 100 °C for $T_{oxd}$ = 2 min oxidized sample grown on glass substrate. Significant optical switching is found in LWIR. Drude-Lorentz oscillator fit to the data is also shown. (b, c) Extracted $VO_2$ complex refractive indexes (n,k) in the insulating and the metallic phase used to model the reflectance data in (a) above. Comparison of the extracted (n,k) with published work [45] is also shown.

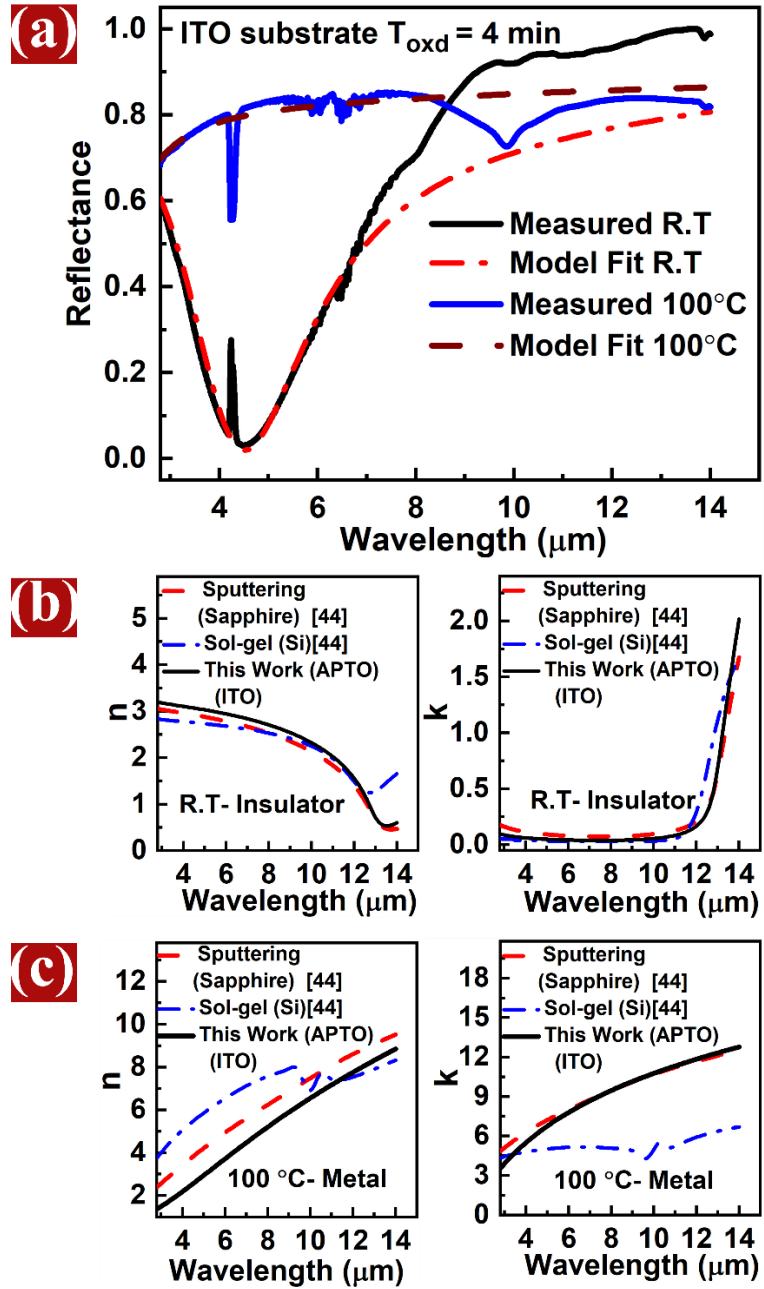

FIG. 4.(a) Measured infrared reflectance spectrum at room temperature and 100 ºC for $T_{oxd}$ = 4 min oxidized sample grown on ITO coated glass substrate. Significant optical switching is found in MWIR. Drude-Lorentz oscillator fit to the data is also shown. (b, c) Extracted $VO_2$ complex refractive indexes (n,k) in the insulating and the metallic phase used to model the reflectance data in (a) above. Comparison of the extracted (n,k) with published work [45] is also shown.